\documentclass{moriond}

\usepackage{amsmath,amssymb}

\def\Journal#1#2#3#4{{#1} {\bf #2}, #3 (#4)}

\def\PRD{{\em Phys. Rev.} D}

\def\be{\begin{equation}}
\def\ee{\end{equation}}
\def\bea{\begin{eqnarray}}
\def\eea{\end{eqnarray}}

\newcommand{\Photo}{}

\begin{document}
\vspace*{4cm}
\title{Modified gravitational wave propagation: information from strongly lensed binaries and the BNS mass function}

\author{ Francesco Iacovelli \footnote{\href{mailto:Francesco.Iacovelli@unige.ch}{Francesco.Iacovelli@unige.ch}\\ To appear in the proceedings of the 56\textsuperscript{th} Rencontres de Moriond 2022, Gravitation session.}, Andreas Finke, Stefano Foffa, Michele Maggiore, Michele Mancarella }

\address{D\'epartement de Physique Th\'eorique and Center for Astroparticle Physics, \\ Universit\'e de Gen\`eve, 24 quai Ansermet,
CH--1211 Gen\`eve 4, Switzerland}

\maketitle\abstracts{
Modified gravitational wave propagation is a smoking gun of modifications of gravity at cosmological scales, and can be the most promising observable for testing such theories. The observation of gravitational waves (GW) in recent years has allowed us to start probing this effect, and here we briefly review two promising ways of testing it. We will show that, already with the current network of detectors, it is possible to reach an interesting accuracy in the estimation of the $\Xi_0$ parameter (that characterizes modified gravitational wave propagation, with $\Xi_{0, {\rm GR}} = 1$) and with next generation facilities, such as the Einstein Telescope, we can get a sub-percent measurement.   
}

\section{Introduction}
The window gravitational waves (GWs) opened on our Universe can lead to outstanding discoveries both in Fundamental Physics and Cosmology. In this context, modified GW propagation has been understood as a particularly interesting property, characterizing all theories that modify General Relativity (GR) on cosmological scales. The best way to probe this effect is to use the so-called ‘‘standard sirens'', i.e. GWs with an associated electromagnetic (EM) counterpart, such as GW170817. Unfortunately, a counterpart measurement is expected only for a small fraction of the GW signals, and more likely at small redshift, where the imprint left by modified gravity is smaller. Other ways of extracting every ounce of information encoded in all detected signals without a counterpart, the so-called ‘‘dark sirens'', turn out to be crucial. The paper is based on the discussion in \cite{ourLens,ourBNS} and organised as follows: in sect. \ref{sec:modGWprop} we review the phenomenon of modified GW propagation; in sect. \ref{sec:quadlens}  a way of extracting information from quadruply lensed binries is presented; in sect. \ref{sec:BNSmfun} we discuss a statistical method based on the BNS mass function.

\section{Modified gravitational wave propagation}\label{sec:modGWprop}
The phenomenon of modified GW propagation arises from a modification of the ‘‘friction term'' in the propagation equation of tensor perturbations (i.e. GWs) over a Friedmann-Lema\^{i}tre-Robertson-Walker (FLRW) background
\begin{equation}\label{eq:prop_eq_MG}
    \tilde{h}_A'' + 2 \mathcal{H} [1 - \delta(\eta)] \tilde{h}_A' + k^2c^2\tilde{h}_A = 0
\end{equation}
where GR is recovered for $\delta(\eta) = 0$. A detailed review is present in \cite{PhysRevD.97.104066,Belgacem_2019JCAP} . The net effect of Eq. \ref{eq:prop_eq_MG} is that, propagating across cosmological distances, the attenuation of the amplitude of GW signals is different with respect to the $\propto 1/a$ behavior of GR. This means that what we measure through GWs is not the ‘‘standard'' luminosity distance [which, following the literature, we refer to as ‘‘electromagnetic'' and denote by $d_L^{\rm em}(z)$], but a ‘‘GW luminosity distance'', which can be parameterised for most of the best studied models as \cite{PhysRevD.97.104066}
\begin{equation}\label{eq:Xiofz_def}
    \frac{d_L^{\rm gw}(z)}{d_L^{\rm em}(z)} \equiv \Xi (z) = \Xi_0 + \frac{1 - \Xi_0}{(1+z)^n}.
\end{equation}
where $\Xi_0=1$ in GR. Given that deviations from GR and $\Lambda$CDM, are constrained to percent level for the background evolution and scalar perturbations by EM probes, such as CMB, SNe Ia, BAO and LSS (e.g. deviations from $w_0 = 1$ are bounded at $3\%$ by \textit{Planck 2018}+BAO+SNe~\cite{2020A&A...641A...6P}), one could naively expect that, even in the tensor sector, the deviations from $\Lambda$CDM will be at most of the same order. In contrast,for viable modified gravity (MG) models, it has been shown that high values, such as $\Xi_0=1.8$, are allowed, meaning that deviations can be as big as 80\%, and the resulting effect can be much bigger and easier to observe than a $w_0$ which is 3\% different from -1. A first estimate of $\Xi_0$ can be extracted from the results obtained using GW170817 \cite{PhysRevD.98.023510} and its counterpart, from which one gets $\Xi_0 \lesssim 14$. More recent and stringent results have been obtained by our group, using statistical methods: from the correlation with galaxy catalogs \cite{Finke_2021GalCat} we find $\Xi_0 = 2.1^{+3.2}_{-1.2}$, and exploiting the BBH mass distribution \cite{Mancarella_2021BBH}, we find $\Xi_0 = 1.2 \pm 0.7$ (68\% c.l.).

\section{Modified GW propagation and quadruply lensed events}\label{sec:quadlens}
\begin{figure}[t]
\vspace{-1.cm}
\hspace{-.4cm}
\begin{minipage}{0.5\linewidth}
\centerline{\includegraphics[width=1.\linewidth]{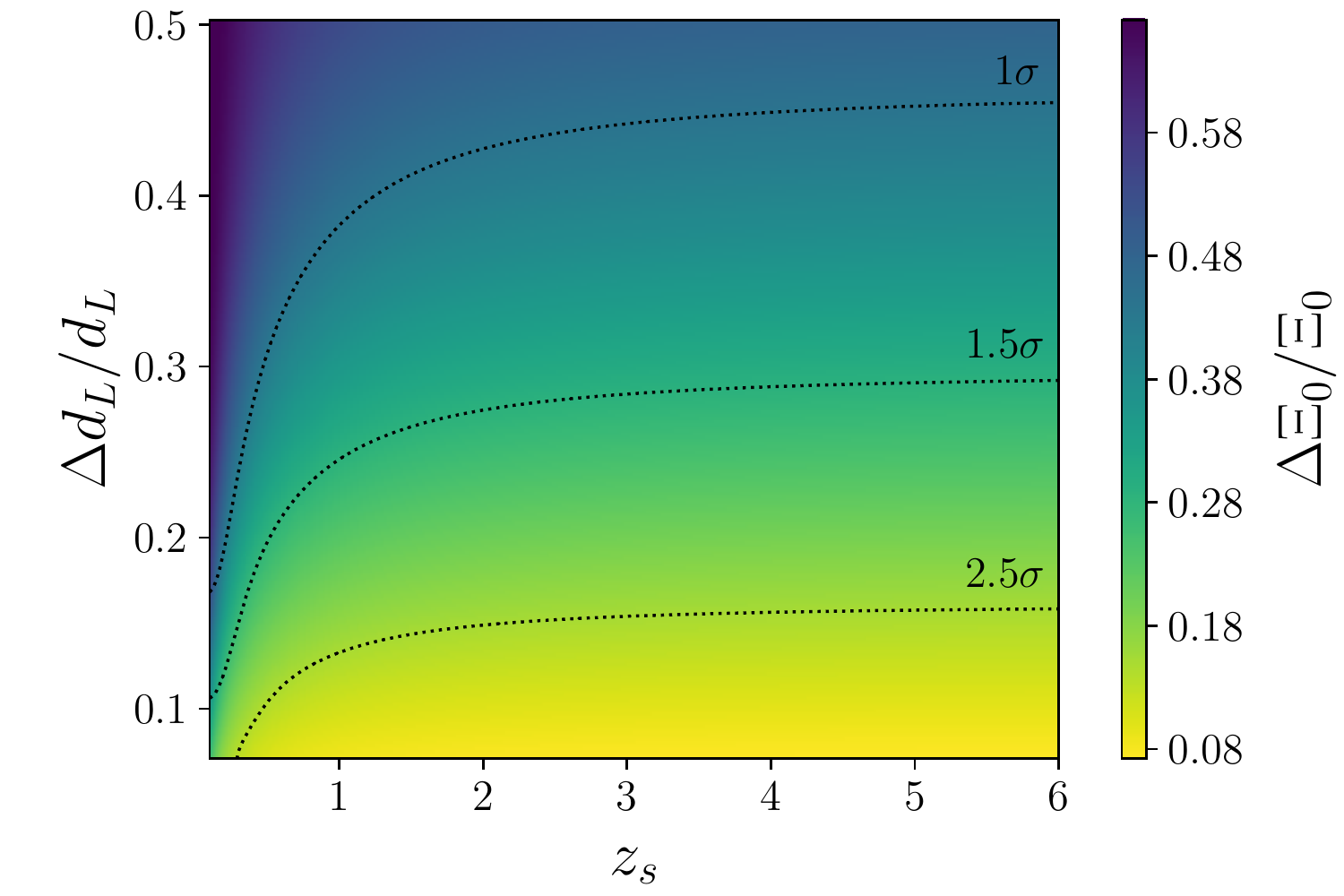}}
\caption[]{Precision attainable on $\Xi_0$ (assuming $\Xi_0=1.8$) from quadruply lensed events, as a function of the redshift and of the combined relative error on the $d_L$ measurements. The lines set the limit for the $1\sigma, \ 1.5\sigma$ and $2.5\sigma$ exclusion of $\Xi_{0,\rm GR} = 1$. Adapted from \cite{ourLens}.}\label{fig:lensing_result}
\end{minipage}
\hspace{.6cm}
\begin{minipage}{0.5\linewidth}
\centerline{\includegraphics[width=1.\linewidth]{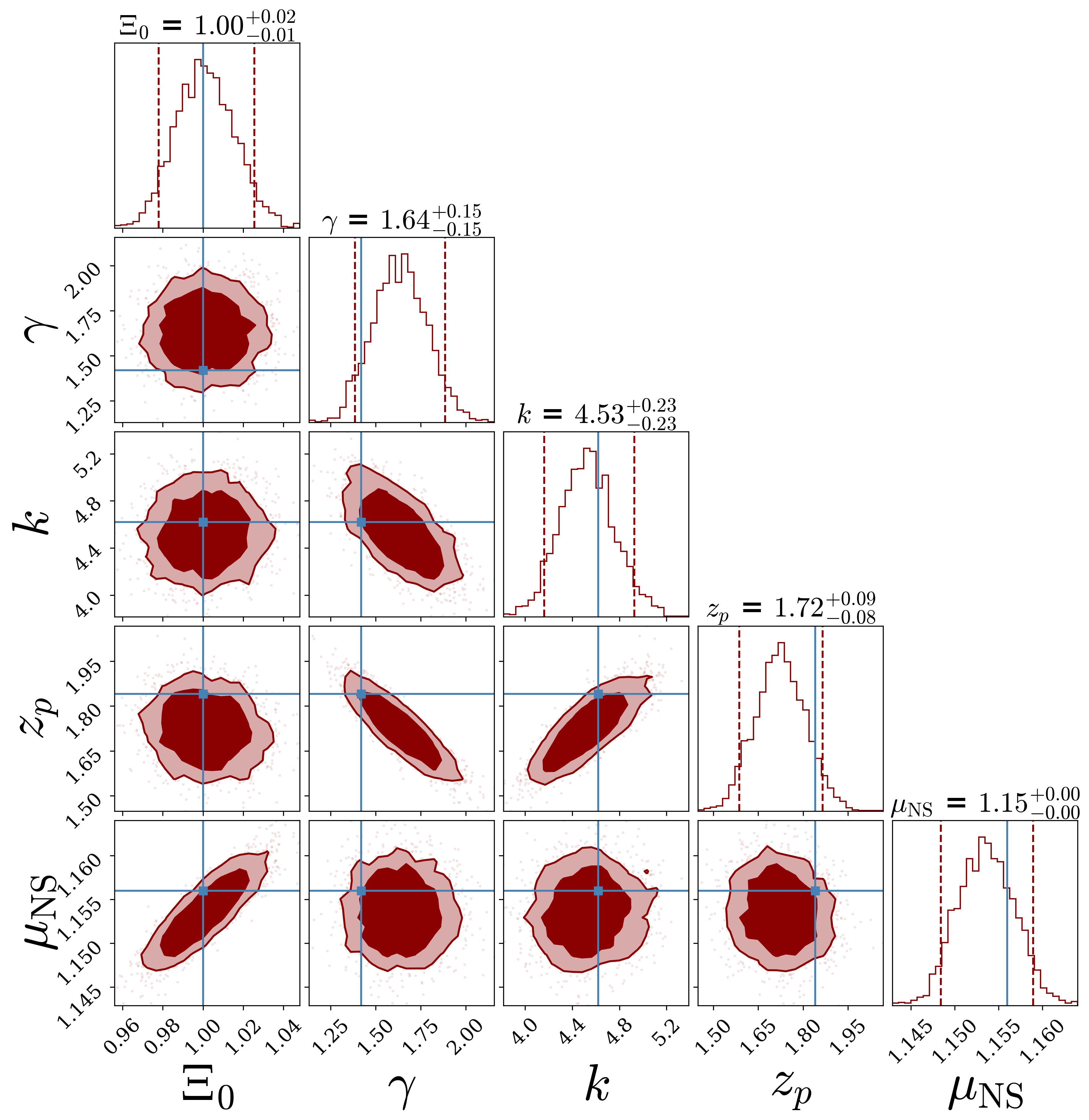}}
\caption{Reconstructed values of $\Xi_0$, the Madau--Dickinson parameters and the mean of the mass distribution, with the method described in sect. \ref{sec:BNSmfun}.}\label{fig:BNS_corner}
\end{minipage}
\vspace{-.5cm}
\end{figure}

GWs can be lensed as any other signal when traveling across cosmological distances, but the angular resolutions of GW detectors is not enough to spatially separate typical strongly lensed images. Signals from coalescing binaries, however, have a short duration, thus strong lensing will manifest itself as a series of repeated GW detections, separated by relative time delays of orders of minutes to months for lensing by galaxies and up to years for lensing by galaxy clusters. Moreover, the amplitude of the various images will differ, since they experience different amounts of magnification or demagnification, but all the other parameters, such as the spins, chirp mass, sky locations, etc., are the same. This allow to perform a Bayesian analysis, and find out whether two or more GW signals belong to the same source \cite{Hannuksela_2019,10.1093/mnras/staa1430}. In particular, in the case of a quadruply lensed event, having the four signals, it is possible to extract \cite{ourLens,Hannuksela_2019} a measurement of the distance from the time delays, $D_{\Delta t}$ and four measurements from the amplitudes, which can be used to perform a cosmological analysis. In particular, it can be shown \cite{ourLens,Tasinato_2021} that the former is not affected by the phenomenon of modified GW propagation, while the latter are, according to Eq. \ref{eq:Xiofz_def}, so their combination can be directly used to infer $\Xi(z)$ without imposing any prior on $H_0$. Assuming that Nature is described by a MG model with $\Xi_0 = 1.8$, the relative error $\Delta\Xi_0/\Xi_0$ that can be obtained from a single quadruply lensed event, as a function of the source redshift and the sum in quadrature of the errors on the luminosity distance coming from the four $d_L^{\rm gw}$ measurements and the time delay one, is presented in Fig. \ref{fig:lensing_result}, for a redshift range appropriate for both LIGO/Virgo/KAGRA and 3G detectors, such as the Einstein Telescope (ET). 

\section{Modified GW propagation and the BNS mass function}\label{sec:BNSmfun}
Together with the luminosity distance to the source, a GW detector also measures the redshifted chirp mass of the binary, i.e. $\mathcal{M}_c = M_c (1+z)$, where $M_c$ is the source--frame chirp mass. Assuming a cosmological model, it is possible to convert the $d_L^{\rm gw}$ measurement into a redshift, to extract $M_c$ but, if Nature is described by a MG model with $\Xi_0\neq 1$, assuming GR and $\Lambda$CDM, one would get a reconstructed redshift, $z_{\rm GR}$, different from the true one, $z_{\rm true}$, and consequently a bias in the estimation of the source--frame mass \cite{ourBNS}. Given the expected narrowness of the NS mass function, which is also predicted to not evolve significantly with redshift \cite{Farrow_2019}, and the impressive detection rate of BNSs achievable at ET and Cosmic Explorer, CE (of order $7\times10^4$ ev/yr already for ET alone \cite{Belgacem_2019GWGRB}), with the possibility to detect these systems up to $z\sim2-3$ for ET \cite{ET_SC} and even $z\sim10$ for CE \cite{Hall_2019} (where the effects of modified GW propagation are huge) among them, within GR, there could not be a single neutron star with ‘‘standard'' mass at large redshift, for high values of $\Xi_0$, like $\Xi_0=1.8$. \cite{ourBNS} Opening up the parameter space to include modified GW propagation, one can then perform a hierarchical Bayesian analysis, using the BNS mass function as a prior, to get an estimate of $\Xi_0$, in the same spirit of what it has been done in the literature for $H_0$. ~\cite{Taylor_2012} The posterior for the desired cosmological and population parameters, marginalised over the overall rate, takes the form described in \cite{Mandel_2019,Fishbach_2018}. The main source of error in this case is the observational error on $d_L^{\rm gw}$, since the relative error on the redshifted chirp mass is of order $\Delta\mathcal{M}_c/\mathcal{M}_c \sim 1/\mathcal{N}_{\rm cyc}$, where $\mathcal{N}_{\rm cyc}$ is the number of observed inspiral cycles, expected to be $\mathcal{O}(10^5)$. The GW likelihood can then be simply approximated as a Gaussian distribution, with $\Delta d_L/d_L = 2/{\rm SNR}$ (the factor of 2 is added to account for the marginalisation over the inclination angle) and we also include the error due to lensing as $(\Delta d_L/d_L)_{\rm lensing} = 0.05 z$, following \cite{Belgacem_2019GWGRB}. For the merger rate, we use a Madau--Dickinson profile with typical parameters \cite{Madau_2017}, and include a time delay distribution $P(t_d)\propto1/t_d$ with $t_{d, \rm min} = 20 {\rm Myr}$. We further adopt a Gaussian distribution for the source--frame chirp mass, with $\mu_{\rm NS}=1.156{\rm M}_{\odot}$ and $\sigma_{\rm NS}=0.056$, obtained from two independent Gaussians for the single masses with mean $1.33 {\rm M}_{\odot}$ and standard deviation 0.09. \cite{Farrow_2019} We perform our analysis on a single ET detector, with triangular shape, adopting the public ET--D sensitivity curve \footnote{\href{http://www.et-gw.eu/index.php/etsensitivities}{http://www.et-gw.eu/index.php/etsensitivities}}, and use only events with ${\rm SNR} > 12$. The computation of the selection effects induced by this choice, fundamental ingredient of this formalism, is performed using a weighted Monte Carlo integration \cite{Tiwari_2018}, which allows to sensibly speed up the computation, essential when using the large catalogs needed to forecast the capabilities of 3G detectors. The posterior distribution for a set of parameters can be estimated with a MCMC sampling. Using these prescriptions, we find that a percent measurement of $\Xi_0$ could be obtained already with $\sim 4600$ well localised events, as shown in Fig. \ref{fig:BNS_corner}, where we used $\Xi_0=1$, while also estimating the mass and rate distribution parameters with high accuracy. 

\section{Summary and conclusion}
Modified GW propagation is one of the most promising observables of MG on cosmological scales. As we showed in sect. \ref{sec:quadlens}, in particular in Fig. \ref{fig:lensing_result}, with a single quadruply lensed GW event, the accuracy we can get in the estimation of $\Xi_0$ is very interesting already with the current network of detectors. Moreover, the error is expected to scale as $1/\sqrt{N}$, with $N$ being the number of events, and at ET we could detect $\sim 25$ quadruply lensed events in 4 years \cite{10.1093/mnras/sty411}. Another promising technique is the one presented in sect. \ref{sec:BNSmfun} since, as we showed, MG can leave a clear signature on GW events, especially at the resdhifts probed by 3G detectors. From Fig. \ref{fig:BNS_corner} we see that, with only 4600 events, the accuracy we get on $\Xi_0$ reaches the percent level, and the posteriors are nearly Gaussian. Also in this case we then expect the error to scale as $1/\sqrt{N}$, meaning that we could get a sub-percent measurement with less than 1 year of ET data.

\section*{Acknowledgments}

The work of the authors is supported by the Swiss National Science Foundation and by the SwissMap National Center for Competence in Research.

\end{document}